# Flexible parametrization of graph-theoretical features from individual-specific networks for prediction

Mariella Gregorich[1], Sean L. Simpson[2] and Georg Heinze[1]

[1]Medical University of Vienna, Center for Medical Data Science, Institute of Clinical Biometrics, Vienna, Austria

[2]Department of Biostatistics and Data Science, Wake Forest University School of Medicine, Winston-Salem, NC, USA

Corresponding author: georg.heinze@meduniwien.ac.at

## Abstract

Statistical techniques are needed to analyse data structures with complex dependencies such that clinically useful information can be extracted. Individual-specific networks, which capture dependencies in complex biological systems, are often summarized by graph-theoretical features. These features, which lend themselves to outcome modelling, can be subject to high variability due to arbitrary decisions in network inference and noise. Correlation-based adjacency matrices often need to be sparsified before meaningful graph-theoretical features can be extracted, requiring the data analysts to determine an optimal threshold.. To address this issue, we propose to incorporate a flexible weighting function over the full range of possible thresholds to capture the variability of graph-theoretical features over the threshold domain. The potential of this approach, which extends concepts from functional data analysis to a graph-theoretical setting, is explored in a plasmode simulation study using real functional magnetic resonance imaging (fMRI) data from the Autism Brain Imaging Data Exchange (ABIDE) Preprocessed initiative. The simulations show that our modelling approach yields accurate estimates of the functional form of the weight function, improves inference efficiency, and achieves a comparable or reduced root mean square prediction error compared to competitor modelling approaches. This assertion holds true in settings where both complex functional forms underlie the outcome-generating process and a universal threshold value is employed. We demonstrate the practical utility of our approach by using resting-state fMRI data to predict biological age in children. Our study establishes the flexible modelling approach as a statistically principled, serious competitor to ad-hoc methods with superior performance.

# 1. Introduction

In recent years, network-based studies have gained popularity, particularly those focused on individual-specific network inference, which involves estimating a separate connectivity matrix (adjacency matrix) across a common set of nodes for each individual subject in the study cohort. A key clinical challenge is to translate the heterogeneous dependency structure among individuals into clinically useful information that can aid in better clinical decision making at the individual level. For instance, functional connectivity in resting-state functional magnetic resonance imaging (rs-fMRI) can be inferred through the correlations of the temporal pairwise blood oxygen level dependent (BOLD) signals among different regions of interest (ROIs) in the brain. However, individual-specific network inference based on observational data can be difficult due to nuisance or confounding factors, which must be taken into account when using these networks to predict disease development or deterioration.

Despite the increasing use and understanding of graph theory, many network inference methods include sparsification as a pre-processing step, which selects the network representative from a large space of possible network representations for an individual. Sparsification procedures are commonly applied to remove weaker links, which are most affected by experimental noise (e.g. physiological processes, subject motion or fluctuations caused by the scanner) not related to the underlying neural activity being studied and thus deemed spurious. The lack of consensus regarding the thresholding strategy keeps this topic an area of ongoing research [1-5]. Currently, there is no commonly agreed upon way to optimally select a threshold, and often ad-hoc modelling approaches are applied. Given a correlation-based adjacency matrix, thresholding is generally performed repeatedly over a sequence of sparsity parameters, and then the properties of the resulting graphs are estimated at each of these threshold values. Instead of selecting a single, optimal parameter value, we propose to use the full sequence of graph-theoretical features obtained at each threshold value and incorporate a flexible weighting function that assigns weights with respect to the relative leverage the property at the threshold has on the outcome. It generalizes the concept of sparsification in individual-specific network inference to a functional setting and offers a novel functional thresholding approach to tackle the uncertainty associated with threshold selection for achieving both adequate network sparsity and good predictive performance.

Motivated by the limited research in individual-specific network sparsification for prediction modelling, we compare the proposed flexible approach with two ad-hoc competitors, optimized threshold selection and averaging over a range of thresholds. This study is intended to encompass an early phase of methodological evaluation (see the concept of phases of methodological research [6]). The remainder of this paper is structured as follows. In Section 2, we introduce the general terminology, followed by a delineation of the current common practice in network sparsification, and our proposed flexible approach. Section 3 presents a comparison of our approach with alternative methods through numerical studies. A real-life data application employing rs-fMRI data of children recorded while watching a movie to predict their age is presented in Section 4. Finally, in Section 5, we discuss the findings. Due to space limitations, a detailed description of the simulation studies and the analysis of the real data example have been relegated to the supplementary material.

## 2. Methods

In this section, we will begin by introducing some general notation and briefly present current statistical practices of using individual-specific networks for prediction. We will subsequently propose a novel flexible approach.

### 2.1. Concepts of graph theory

Given relational data per individual $i, i = 1, \ldots, N$, the complex connection pattern can be represented by a network where, for each individual, a separate connectivity matrix (adjacency matrix) is estimated over a common set of nodes. In the field of neuroscience, these nodes may correspond to functional areas of interest within the brain's connectome, where the strength of their connections (edge weight) reflects the temporal correlation of BOLD signals. For purpose of analysis, an undirected network can be mathematically represented by a symmetric adjacency matrix $A_i \in \mathbb{R}^{p \times p}$. In this matrix, the entries $a_i(r,s)$ represent the connection strength between the nodes $r$ and $s$ in individual $i$, and hence an $a_i(r,s) \neq 0$ indicates that nodes $r$ and $s$ are connected, and $a_i(r,s) = 0$ indicates that they are not connected. The connection strengths could be estimated by various methods, e.g. marginal correlation between time series of signals, partial correlation, mutual information, etc. For the sake of simplicity, we have chosen to use marginal correlation throughout this work, which is a common choice in network analysis. When correlation-based estimates of connectedness are based on empirical data, the case that any edge weight $a_i(r,s) = 0$ will occur, yielding a fully connected network. Without loss of generality, we assume $a_i(r,s) \in [0,1]$ for any $i, r, s$ as any interval can be transformed to this range (e.g. setting negative values to zero). Sparsifying the initial adjacency matrix means removing weak connections such that only a proportion of all pairs of nodes are connected. Graph theory then provides the basis to describe the network's properties using graph-theoretical features. For example, the network could be described by the characteristic path length (CPL) or by its clustering coefficient (CC). Finally, such graph-theoretical features lend themselves as predictors in a regression model to predict a trait $y$ (Figure 1).

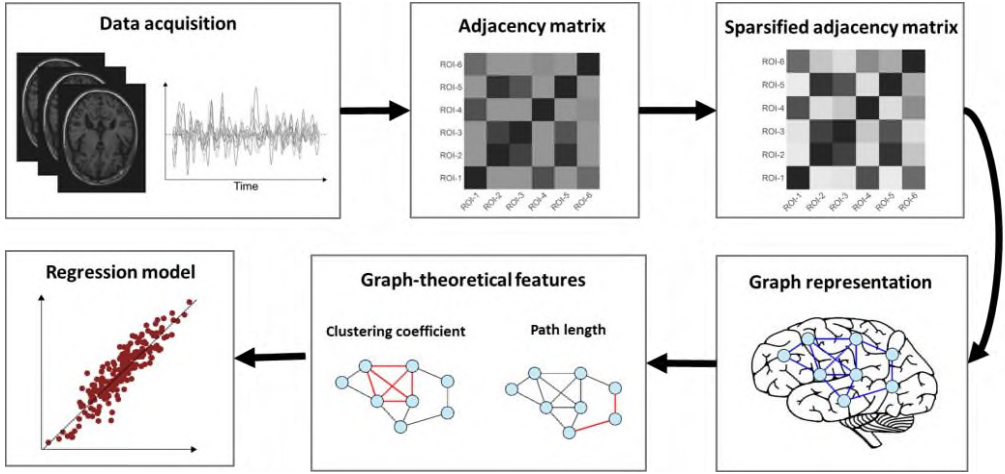

*Figure 1. Process of individual-specific network inference for prediction modelling*

With most methods of sparsification, network inference depends on a parameter $t$, which is the threshold that is applied to remove apparently weak or spurious connections. Commonly, it is either a maximum value for the density (density-based thresholding) or a minimum edge weight (weight-based thresholding) [7]. Approaches to sparsify the individual-specific networks are not tied to the specific method used to estimate the connection strength. Neither for density-based nor for weight-based thresholding is there a commonly accepted consensus on the exact threshold that should be used to infer the individual-specific networks $A_i$, and various methods are used in practice to determine the threshold at which graph-theoretical characteristics become relevant for prediction.

## 2.2. Prediction modelling with sparse networks

In this paper, we consider sparsification as a function of the threshold. Let $x_i(t)$ denote a graph-theoretical feature that describes the network of individual $i$ obtained from applying threshold $t$ to the initial adjacency matrix $A_i$ of subject $i$, $i = 1, \ldots, N$. We will denote the theory for a single graph-theoretical feature, but its application can be extended to multiple features. We now assume that the outcome variable is regressed on a weighted sum of a sequence of such features resulting when applying a sequence of thresholds $T \in [0,1]$ to the graphs. We assume a continuous outcome variable in a linear regression model

$$y_i = \beta_0 + \beta_1 \sum_{t \in T} \omega(t) x_i(t) + \sum_{j=2}^{k} \beta_j z_{ij} + \epsilon_i, \quad i = 1, \ldots, N \qquad (1)$$

where $z_{ij}$, $j = 2, \ldots, k$ are some auxiliary covariates of subject $i$ and $\epsilon_i$ is some error term. Previous attempts to exploit the information in the sequence of features $x_i(t)$ available for a subject $i$ concentrate mainly on two approaches. Both approaches rely on a restricted predefined subsequence $\tilde{T} \subseteq T$.

Optimal threshold approach (OPT): the first approach presupposes the existence of an optimal universal threshold $t_{opt}$ such that

$$\omega(t) = \begin{cases} 1, & \text{for } t = t_{opt} \\ 0, & \text{for } t \neq t_{opt} \end{cases}$$

To find the optimal threshold, the sum of squares of $\epsilon_i$ (or more generally, a loss function $f_{loss}$ applied to the predictions $\hat{y}$ and observed outcomes $y$) is minimized over the sequence of models defined by $t \in \tilde{T}$, such that $t_{opt} := \underset{t \in \tilde{T}}{\operatorname{argmin}} f_{loss}(y, \hat{y})$. In case of a correlation-derived adjacency matrix, the search is commonly performed in $\tilde{T} \in [0.1, 0.4]$ [7].

Average feature approach (AVG): in the second modelling approach, the weight function is considered to be piecewise-constant across a specified sequence $t \in \tilde{T}$ such that

$$\omega(t) = \begin{cases} \frac{1}{|\tilde{T}|} & \text{, for } t \in \tilde{T} \\ 0 & \text{, for } t \notin \tilde{T} \end{cases}$$

where $|A|$ denotes the cardinality of the set $A$. Similarly to the OPT approach, the respective subset is $\tilde{T} \in [0.1, 0.4]$ in correlation-based network inference, with the difference that all resulting graph-theoretical features which result from the application of the equidistant thresholds in the subset are averaged. In practice, this is supposed to avoid choosing a non-optimal threshold [7].

Flexible weighting approach (FLEX): the weight function is estimated through flexible functions such that

$$\omega(t) = \sum_{m=1}^{M} \theta_m b_m(t) \tag{2}$$

where $b_m(\cdot)$ denotes a set of $M$ functional basis function of $t$ and $\theta_m$ the corresponding weight, $m = 1, \ldots, M$. The functional form of the weight function $\omega(t)$ is unknown and its shape has to be estimated. Functional basis forms can be defined by restricted cubic splines, B-splines or penalized splines [8], while the effect magnitude can be estimated together as

$$\gamma = \beta_1 \theta_m, \quad j = 1, \ldots, m$$

In the presence of prior knowledge about a plausible shape of the weight function, outer knots could be placed such that e.g. $\omega(0) = 0$, while interior knots are placed in equidistant steps.

FLEX strives to learn the most suitable functional form $\hat{\omega}(t)$ for the full sequence of thresholds $T$ based on the available data instead of assuming a universally optimal threshold or treating a subset of thresholds as equally appropriate. Knowing where $\hat{\omega}(t)$ assumes large or small values provides information about at which thresholds $t$ values of $x(t)$ will have the greatest leverage on the predictions $\hat{y}$, and may also help to support the assumptions made in the OPT or AVG approaches, or to define a suitable subsequence $\tilde{T}$. Such information can also be very useful for choosing appropriate sparsification thresholds if the individual networks are not only used for prediction but their structures and topologies are also interpreted individually. The formulations above are not entirely new and similar concepts have been used for modelling time-dependent exposures [9] or in functional data analysis [10].

## 3. Simulation studies

### 3.1. Aim

We conducted a plasmode simulation study [11] in which we created data closely resembling the complex network structures of real fMRI data but with additional features and corresponding outcomes. The aim of the simulation study was (i) to provide evidence for a proof-of-concept of our proposed methodology, and to (ii) evaluate its predictive performance comparing it to common ad-hoc modelling approaches.

### 3.2. Motivating example and data-generating mechanism

We used a subset of the publicly available data in the pre-processed repository from the ABIDE initiative, which is a multi-site observational data collection of functional and structural brain imaging data on autism spectrum disorder (ASD) [12,13]. The ABIDE initiative aims to understand the extent to which ASD impacts brain functional connectivity at rest and to investigate how age interacts with the effect of ASD on resting-state functional connectivity. The functional pre-processed data of the ABIDE Preprocessed initiative comprises 1112 observations from 539 individuals with Autism spectrum disorder (ASD) and 573 age-matched typical controls (TC) from seventeen sites [13]. The full cohort was restricted according to the quality assurance and sample selection specified in Di Martino et al. [12], which includes, inter alia, the restriction to only male participants due to the low number of females (N=164), yielding N=704. In addition, we omitted 18 observations with missing or unmeasured time series, leaving N=686 individuals with corresponding pre-processed fMRI data. In practical applications where multiple cohorts contribute to a study, center effects may have to be considered to adjust for different fMRI pre-processing steps with approaches like ComBat [14,15] but we don't assume them to be present here.

We computed $P \times P$-dimensional adjacency matrices $A_i, i = 1, ..., N$, based on Spearman correlations between the time series (P=116) for each individual. Each individual-specific network consisted of 116 nodes. Details of the fMRI data pre-processing can be found in the Supplement Section 1.1. We varied the sample size $N = \{75, 150, 300, 600\}$ by randomly selecting subsets without replacement. For feature extraction, the two most frequently used graph-theoretical features, the clustering coefficient and the characteristic path length, were computed, which served as our true predictor $x(t)$ evaluated at weight-based thresholds $t \in T$. Supplementary Figure 2 illustrates the variations in graph-theoretical features across weight-based thresholds. We focused on these two features because of their frequency of use and their distinct patterns of curve shapes, location of peaks, and areas of increased variability across thresholds (see Supplementary Figure 2). Our outcome-generating mechanism (OGM) assumed the following form

$$E[Y|x(t)] = \beta_0 + \beta_1 x^*$$

with $x^*$ defined as either

- *universal* threshold: $x^* = x(\tau)$ with $\tau = 0.25$
- *random* threshold: $x_i^* = x_i(\tau_i)$ with $\tau_i \in U(0.1, 0.4)$
- *functional* threshold: $x_i^* = \sum_{t \in T} \omega(t) x_i(t) \; with$
    I. *flat* functional form: $\omega(t) = 4$
    II. *early peak* functional form: $\omega(t) = 6.5\sin(2\pi t)I(t < 0.5)$
    III. *arc* functional form: $\omega(t) = 6\sin(\pi t)$

and drawing outcomes $y_i = E[Y|x_i(t)] + \epsilon_i$ with Gaussian errors $\epsilon \sim N(0, \sigma^2)$ accordingly. Illustrations for the universal and random threshold selection are depicted in Supplementary Figure 3, while functional forms are shown in Supplementary Figure 4.

Residual variance:

The variance $\sigma^2$ of the residuals $\epsilon$ was specified such that in each of the OGMs, the true model (oracle) yields an $R^2$ of 1, 0.6, or 0.3 corresponding to no, moderate, and high residual variance, respectively, and was determined in a pilot study prior to the simulation study. The scenarios with $R^2 = 1$ may seem unrealistic but were included as a benchmark to study bias resulting from model and/or network misspecification. The values of $\sigma^2$ are shown in Supplementary Tables 2.

Contamination of the edge weights:

In applied network inference, the edge weights in $A_i$ cannot be reliably estimated. This was reflected in our simulation by introducing contamination to the originally measured edge weights $a_i(r, s), \forall (r, s)$ [16]. More specifically, we added stochastic perturbations to the correlation-based adjacency entries for each individual to create contaminated adjacency matrices $\tilde{A}_i = g(A_i, u_i, \alpha)$ according to a modified version of the strategy proposed by Hardin et al. [16]. While full details are given in the Supplement, here we only stress that the parameter $\alpha$ controls the maximum allowed deviation of any edge weights between the original and contaminated adjacency matrices. Similar to the residual variance, three levels of contamination were examined: none ($\alpha = 0$), moderate ($\alpha = 0.15$) and high contamination ($\alpha = 0.3$).

For each combination of the parameter values, i.e. magnitude of residual variance, degree of contamination and sample size, the graph-theoretical features (clustering coefficient, CC; characteristic path length, CPL) and the five OGMs (*universal*, *random*, *flat*, *half-arc* and *arc*), we simulated 500 datasets (as a balanced compromise between precision and managing computational runtime, see Supplement Section 1.2.3), each comprising $N$ individual-specific adjacency matrices $\tilde{A}_i$ and the outcome $y_i$. Note that outcomes were generated before perturbing the adjacency matrices.

### 3.3. Estimands and targets

The estimands in our simulation were the expected values of the outcome variable $E(Y|A_i)$ and the weight functions $\omega(t)$.

### 3.4. Methods of data analysis

Given the (contaminated) adjacency matrices $\tilde{A}_i$, the discrete sequence of (contaminated) graph-theoretical features $\tilde{x}_i(t)$ was computed for $T = \{0, 0.01, 0.02, ..., 1\}$. This was done for the clustering coefficient and the characteristic path length, both under density-based and weight-based thresholding, yielding 4 sequences $\tilde{x}_i(t)$ forming the basis for the following procedures.

The OPT, AVG, and FLEX approach were implemented with varying threshold subsets considering the thresholding strategy (weight- or density-based) and the respective analysis approach (OPT, AVG, FLEX), as summarized in Table 1. The optimal threshold according to the OPT approach was identified by cross-validation with respect to the minimal cross-validated RMSPE. The weight function $\hat{\omega}(t)$ for the sequence $\tilde{x}(t), t \in \tilde{T}$, in FLEX was estimated by penalized splines with a fixed number of segments of 25 and cubic degree. In addition, we implemented two benchmark methods, the NULL and the ORACLE approach. The former served as a lower benchmark and took the average of the observed $y$ as its prediction $\hat{y}$, whereas the latter (upper benchmark) received the true outcome-generating weight function and was only included in case of weight-based analysis, as the true weight function was defined for weight-based thresholding only.

*Table 1. Overview of methods parameters and subset specifications*

| Method | Thresholding | Threshold subset | Threshold(s) selection |
|---|---|---|---|
| **OPT** | Weight-based | $\tilde{T} = \{0, 0.01, ..., 0.75\}$ | Threshold selection by cross-validation within subset $\tilde{T}$ |
| | Density-based | $\tilde{T} = \{0.25, 0.26, ..., 1\}$ | |
| **AVG** | Weight-based | $\tilde{T} = \{0.1, 0.11, ..., 0.4\}$ | Averaging over all features $\tilde{x}(t), t \in \tilde{T}$ |
| | Density-based | $\tilde{T} = \{0.6, 0.61, ..., 0.9\}$ | |
| **FLEX** | Weight-based | $\tilde{T} = \{0, 0.01, 0.02, ..., 1\}$ | Penalized splines with a fixed number of segments of 25 for the weight function $\hat{\omega}(t), t \in \tilde{T}$ |
| | Density-based | $\tilde{T} = \{0, 0.01, 0.02, ..., 1\}$ | |

**3.5. Performance measures**

We evaluated the following performance measures of the prediction models:

- root mean squared prediction error (RMSPE),
- the $R^2$,
- calibration slope (CS),

These performance measures were evaluated in each simulated data set using 5-fold cross-validation (CV), and then averaged over the simulation repetitions of each scenario. The RMSPE values were also normalized by dividing by the minimum RMSPE achieved in a scenario (relative RMSPE) and shown as boxplots. We also report the 2.5% and 97.5% percentiles of the distribution of each performance measure over the simulation repetitions.

Further details of the simulation settings are presented in Supplementary Material 1. All code and implemented functions used to conduct the presented numerical experiments can be found online on GitHub: https://github.com/mgregorich/PRONET.

**3.6. Results**

3.6.1. Instability of the OPT model

In the OGM *universal*, we assumed the functional form to be an indicator function with $\omega(0.25) = 1$ and 0 otherwise to resemble a universal threshold for the entire study cohort creating a scenario favouring the OPT model, whereas approximating an indicator function using penalized splines in this setting might prove difficult. While under close to optimal conditions ($R^2 = 1$ and moderate edge weight contamination), OPT demonstrated on average the best and also most stable prediction performance, in particular when considering CPL and when performing weight-based thresholding (see Supplementary Figure 6, OGM *universal*). However, in a more realistic setting with $R^2$ of 0.3 and moderate edge weight contamination as stated in Table 2, the OPT approach had higher RMSPE than its competitors, in particular compared with AVG (AVG: 5.03, 95% CI 4.29- 5.72; OPT: 5.14, 4.38-5.86). In the latter setting, the AVG method performed similarly to the Oracle model, particularly when considering the clustering coefficient (AVG: 5.03, 95% CI 4.29- 5.72; Oracle: 5.03, 4.30-5.72). Despite the challenges in accurately approximating an indicator function in this particular scenario, FLEX performed slightly better than OPT (FLEX: 5.11, 95% CI 4.38-5.86). Across all OGMS, OPT often underperformed on average compared to its competitors, as shown in Supplementary Figure 6. When OPT was conducted to identify the optimal density level, a low variability of both explanatory features and the additional difficulty introduced by tuning the optimal threshold by cross-validation led to considerable uncertainty in model estimation (see Figure 2).

*Table 2. Simulation study: Average performance measures across simulation replicates for the setting of sample size n=75, moderate residual variance, high level of contamination of the edge weights and the clustering coefficient for OGM universal and weight-based analysis*

| Model | Performance | | | |
|---|---|---|---|---|
| | RMSPE (95% CI) | Relative RMSPE (95% CI) | R2 (95% CI) | CS (95% CI) |
| **Oracle** | 5.027 (4.302, 5.716) | 1.005 (1, 1.028) | 0.296 (0.167, 0.438) | 1.060 (0.927, 1.226) |
| **OPT** | 5.135 (4.375, 5.861) | 1.027 (1, 1.08) | 0.270 (0.139, 0.412) | 1.050 (0.864, 1.314) |
| **AVG** | 5.031 (4.291, 5.715) | 1.006 (1, 1.028) | 0.295 (0.167, 0.436) | 1.062 (0.925, 1.24) |
| **FLEX** | 5.109 (4.376, 5.863) | 1.021 (1, 1.064) | 0.28 (0.146, 0.421) | 0.992 (0.812, 1.188) |
| **Null** | 5.815 (4.989, 6.607) | 1.165 (1.062, 1.313) | - | - |

3.6.2. Gain in performance by flexible modelling

In Figure 2, the relative root mean squared prediction error (RMSPE) of four model-building methods (OPT, AVG, FLEX, and Oracle) was compared for all OGMs, with high contamination of the edge weights. The FLEX model outperformed other methods across all OGMs in the absence of residual variance. With decreasing $R^2$, the performance of the modelling approaches showed little deviations when weight-based thresholding was applied; however, FLEX consistently maintained comparable or marginally superior predictive performance, in particular for more complex functional forms (early peak and arc form). Extreme values in relative RMSPE occurred with the OPT methodology, especially with small samples (n=75, 150) and density-based analysis (see Figure 2 and Supplementary Figure 7). The AVG modelling approach outperformed the OPT method and even demonstrated comparable performance to FLEX with decreasing $R^2$. However, in terms of calibration, the AVG approach exhibited instability when the OGM employed functional forms largely deviating from the intrinsic weighting function of AVG (i.e. *flat* and *arc*) and the characteristic path length was evaluated (*flat*: 0.057, 95% CI -15.34-14.29); *arc*: 12.35, 95% CI -13.40-11.84; refer to Supplementary Table 3). Overall, the FLEX method appeared to be a reliable and robust method for analysing network data. The results provided in Figure 2, based on a sample size of $n = 75$, are consistent with the findings illustrated in Supplementary Figure 8, which demonstrates no variations of the main conclusions when varying sample sizes.

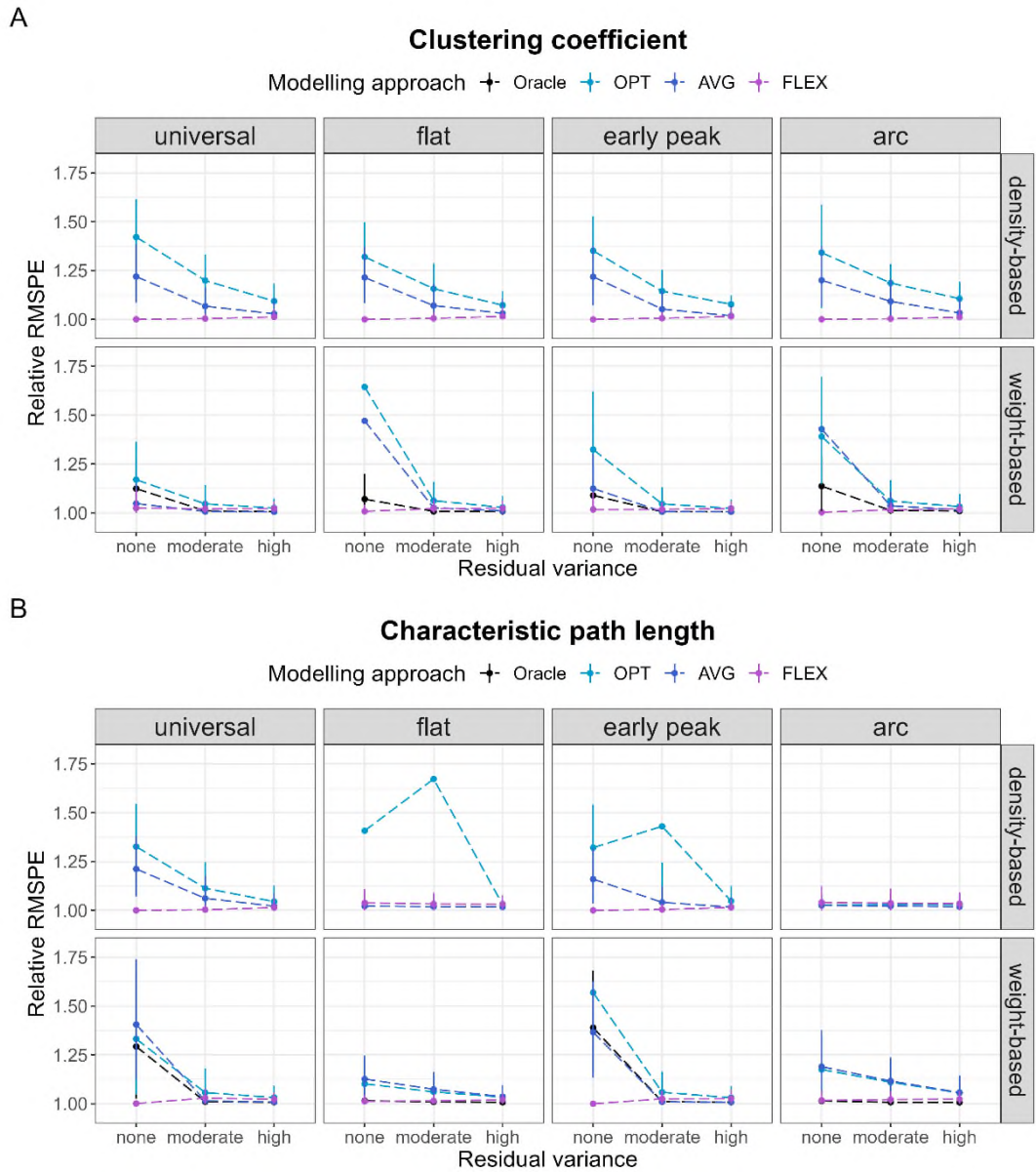

*Figure 2. Simulation study: Relative RMSPE for various outcome-generating mechanisms for the predictive graph-theoretical features (A) clustering coefficient and (B) characteristic path length for a sample size of 75 under density-based and weight-based thresholding, respectively.*

### 3.6.3. Impact of contamination on functional forms and threshold selection

The estimated functional form with 95% confidence intervals of FLEX across different OGMs is presented in Supplementary Figure 9, considering varying levels of edge weight contamination and a fixed sample size of 150. The estimated coefficient functions were obtained by averaging over 500 simulation replicates to derive the average functional form. To compare with the true underlying functional form of each OGM, weight-based thresholding was employed. The FLEX approach demonstrated shape preservation up to moderate levels of edge weight contamination. However, at high levels of contamination, the shape became less discernible in the majority of scenarios.

In Supplementary Figure 10, the accuracy of threshold selection for the model-building approach OPT is depicted across varying sample sizes, levels of residual variance and contamination of edge weights. The red dashed line indicates the true threshold for the OGM *universal*. Despite the clear advantage of the OPT approach in the *universal* OGM, it was apparent that the introduced nuisance factors for the outcome and the edge weights had a notable influence. This was evident when comparing the dashed line to the solid red line. While the variation within the range may not be substantial relative to the y-axis, adjustments in the threshold can have a significant effect on edge weight reduction due to the pronounced right-skewed distributions of edge weights (refer to Supplementary Figure 1). In addition, the OPT approach effectively captured the functional form in the OGM *early peak*, where the coefficient function $\omega(t)$ exhibited its highest leverage point at $t = 0.25$. In the OGM *random*, a threshold was randomly chosen from a uniform distribution $U(0.1, 0.4)$ for each individual, hence, when utilizing the OPT model, the selected threshold should also be centered around 0.25 on average. Supplementary Figure 9 demonstrated that this held true with minimal contamination in both the outcome vector and the edge weights. However, at higher contamination levels, as seen in the 'universal' outcome-generating scenario, there were more significant fluctuations.

In summary, when considering increasing residual variance and contamination of the edge weighting, the FLEX approach exhibited a decline in shape preservation of the true functional form, while OPT demonstrated a relatively robust selection of the underlying threshold despite these nuisance factors. However, in terms of predictive performance, FLEX gained an advantage from its greater adaptability due to its flexibility and, as a result, achieved improved performance compared to the ad-hoc alternatives.

## 4. Real data analysis: predicting the age of children by rs-fMRI

We used data from a study conducted by Richardson et al. [17] with publicly available rs-fMRI data recorded in children while they were viewing a short animated film. The goal of the analysis was to 'predict' the age of the children based on the rs-fMRI measurements. Details of the study design and data pre-processing can be found in the Supplementary Material (section 1.1). The full study cohort comprised 122 children with an average age of 6.7 years (2.3) and 33 adults with an average age of 24.8 years (5.3). We restricted the study cohort to only children. We evaluated the out-of-sample performance of the competitor modelling approaches by internal 5-fold cross-validation (CV) with 10 repetitions. For the OPT approach, optimal threshold search was performed within the interval $\{0, 0.01, ..., 0.75\}$, thus including the option for no thresholding. For the AVG approach, we averaged graph-theoretical features within the interval {0.1, 0.11, ... , 0.4} with equidistant thresholds with a step size of 0.01. For FLEX, we used penalized splines with 25 segments and cubic degree. The results for the graph-theoretical feature characteristic path length are not presented, as there seemed to be no association with age. Gender was included in the model building procedure for adjusted analyses; however, its inclusion did not result in any improvements.

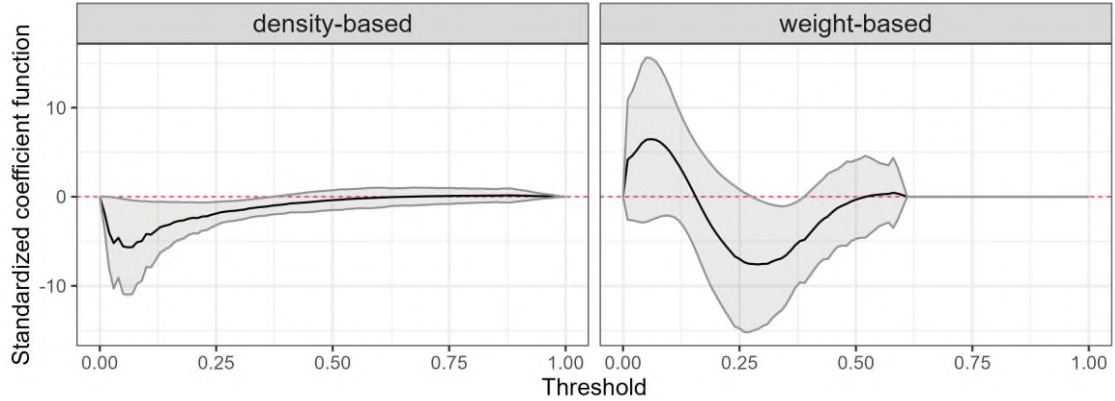

*Figure 3. Real data analysis: Estimated flexible coefficient function $\hat{\beta}(t)$ and pointwise 95% confidence interval for the FLEX model of the association between the clustering coefficient and age. The fit is standardized by the empricial standard deviation of the graph-theoretical feature at each threshold to account for large values at the extremes due to low variability of the feature.*

In contrast to the coefficient function in the density-based thresholding procedure, the functional form for the weight-based thresholding procedure in Figure 3 (right) displays more variability, characterized by fluctuations and shifts in direction as the threshold increases. For the latter, the estimated curve exhibits multiple peaks with the global absolute maximum occurring at $t = 0.29$ with $\hat{\beta}_{std}(0.29) = -7.6$ (95%CI: $-14.8, -0.3$). Intervals of $t$ where $|\beta(t)|$ is large are influential for the prediction of age in children, with the sign implying a negative association for the cc. However, interpretation is difficult given the change in direction of the association and the high correlation of 0.84 between the clustering coefficient for networks thresholded at $t = 0.06$ and $t = 0.29$. The shape of the smooth function indicates the relative leverage of thresholds in density-and weight-based thresholding and the additional importance of the clustering coefficient evaluated for thresholds around 0.29. The selected OPT threshold with a value of $t_{opt} = 0.25$ falls within a comparable range. However, the OPT and AVG modelling approaches did not detect an association with the outcome according to the $R^2$ and the higher RMSPE. When density-based thresholding was conducted, we achieved comparable values of RMSPE, the $R^2$ and the calibration slope for AVG and FLEX as stated in Table 3. The results align with the functional form of the FLEX modelling approach, which assigned fairly homogeneous weights to the thresholds (Figure 3, left), resembling piecewise constant splines. OPT selected the density-based threshold of 0.9 as optimal but did not reach similar predictive performance as its competitors. Further, the weight-based analysis generally suffered more from overfitting (particularly AVG and OPT) compared to the density-based analysis, where the calibration slope was closer to 1. In the density-based approach, FLEX was perfectly calibrated ($CS = 1.00$), whereas AVG slightly underfitted ($CS > 1$).

Table 3. Real data analysis: Predictive performance of the three modelling approaches using weight-based and density-based thresholding. Grid search for optimal threshold selection was only conducted in the OPT approach. RMSPE, root mean squared prediction error, CS, calibration slope

| Method | Thresholding | Threshold | RMSPE | $R^2$ | CS |
|---|---|---|---|---|---|
| FLEX | weight-based | - | 7.707 | 0.134 | 0.861 |
| AVG | weight-based | - | 8.118 | 0.019 | 0.744 |
| OPT | weight-based | 0.25 | 8.148 | 0.019 | 0.488 |
| AVG | density-based | - | 7.669 | 0.126 | 1.114 |
| FLEX | density-based | - | 7.680 | 0.138 | 1.002 |
| OPT | density-based | 0.9 | 7.976 | 0.096 | 0.978 |

The findings of the applied data example exhibit similarities with those of the simulation study, albeit within a diagnostic framework. The explained variability of the graph-theoretical features is less pronounced compared to the simulated dataset, yet it still showcases the added predictive value of individual-specific networks. In cases involving more intricate functional forms, FLEX demonstrates superior performance. Conversely, in scenarios where simpler functional patterns resembling a piecewise-constant weighting are present, FLEX still achieves comparable performance to AVG.

## 5. Discussion

With the increasing availability of relational data, it is becoming more critical to address methodological gaps persisting in network inference. Many neuroimaging studies have demonstrated the potential of functional network connectivity patterns estimated from rs-fMRI to discriminate groups or to predict a clinical outcome [7]. Two modelling approaches are mainly employed in applied studies to address the variability of graph-theoretical features resulting from the selection of thresholds across a broad spectrum of sparsity levels. The first one selects a single threshold based on optimal prediction (OPT), while the other one averages the resulting graph-theoretical feature over a predefined range of thresholds (AVG). We explored an alternative method, referred to as FLEX, which incorporates flexible weighting of the full range of possible thresholds instead of the traditional approach of selecting a single threshold or assigning equal weight to a restricted subset of thresholds.

In the plasmode simulation study, we used the pre-processed real-life fMRI data from the ABIDE initiative ensuring that the complex dependencies inherent in fMRI data were fully conserved and generated a synthetic outcome variable. In addition, the impact of noisily measured networks was assessed by contaminating the edge weights according to a realistic noise model. The FLEX model was able to recover a variety of plausible shapes of the true weight function in the absence and presence of contamination and provided great predictive accuracy and calibration of the predicted outcome, specifically when OGM consisted of complex functional forms. The various levels of contamination demonstrated the FLEX model to be the most resilient to contamination of edge weights and to achieve well-calibrated predictions, suggesting its effectiveness in real-world scenarios, followed by the AVG approach. While the plasmode simulation design was motivated by neurological data settings, the methodology is designed and presented in generality and is applicable to many other areas of scientific research.

In the real data example, we compared the three modelling approaches in their ability to predict age in children based on rs-fMRI data. In weight-based thresholding, however, only FLEX was able to establish an association between the clustering coefficient and the age of the children. In density-based thresholding, we could only see a minimal advantage of FLEX over AVG, probably because the additional parameters that FLEX needs to be estimated are not well supported, and equally weighting the features over the thresholds created more stability. Both approaches outperformed OPT (as in the simulation study), which needs another layer of cross-validation to determine the optimal threshold, and this generates additional estimation uncertainty. A similar effect was noted for regularized regression when the penalty strength has to be estimated from the data [18]. Therefore, when using cross-validation in small data sets to determine the optimal approach (as exemplified in the analysis of the applied data example) one should be aware of these effects and probably prefer stability over flexibility. Both in the simulation study and in the applied example, the OPT approach proved to be unstable and satisfactory performance strongly dependent on sufficient sample size.

In practice, the proposed method can be applied in two settings. First, it can supplement a conventional primary analysis with flexible weights for the full range of possible thresholds by allowing data-driven flexible estimation of threshold leverage and perhaps more stable weighting without restricting the search within a subset of thresholds. By allowing weighting to be determined by predictive leverage, substantial gains were possible compared to the competitor methods, which require an additional layer of cross-validation (usually not performed in applied studies) in the OPT modelling approach, or the prespecification of a subset of thresholds in the AVG approach. Second, the flexible parametrization approach could be the primary analysis, specifically supported in larger sample sizes to warrant the additional complexity, and if more complex functional forms are assumed. Clearly, the FLEX model does not require predefined subsets at the planning stage but additional degrees of freedom due to the flexible parametrization. When comparing the modelling approaches, it was found that OPT exhibited instability, particularly when dealing with small data sets. If the assumption of piecewise-constant weighting is valid, the AVG approach yields robust and accurate prediction results in weight-based analysis. However, to confirm the accuracy of this assumption, additional applied studies need to be conducted to assess the adequacy of the AVG modelling approach. Structures and topologies of the network, as captured by graph-theoretical features, may differ by the threshold applied for sparsification, and hence flexible weighting seems most appropriate to consider their effects on the outcome. In addition, several extensions of our approach are possible. For example, prior knowledge about the possible shape of the functional form can be incorporated to facilitate estimation by imposing a constraint, ensuring it aligns with a predetermined value at threshold t.

We have provided two novel contributions to the assessment of individual-specific networks for prediction modelling. First, we introduced the FLEX approach, which achieved comparable or superior performance as existing approaches, both under contamination of the graphs and regardless of the thresholding technique used. By leveraging graph-theoretical features and considering the dynamic interplay between networks and outcomes, we can gain valuable insights and improve our understanding of complex systems, thereby enhancing diagnostic capabilities in various domains. Second, this study compared several sparsification approaches in a synthetic setting under a variety of realistic data-generating settings using real-life individual-specific networks. Considering the current lack of consensus in the applied literature, our study establishes a solid foundation for the choice of the thresholding approach and serves as a stepping stone for further applied exploration of appropriate modelling techniques in prediction modelling utilizing individual-specific networks. We found that the additional parameters that FLEX needs to estimate need to have support in the data set, and in very small studies the AVG approach might be preferable for its stability. However, both approaches are clearly superior to selecting a single optimal threshold.

## Conflicts of interest

None of the authors have a conflict of interest to disclose.

## Funding

Mariella Gregorich received funding for this work as part of the DOC fellowship from the Austrian Academy of Sciences.

## Data sharing

The ABIDE data set[12,13] is freely available at http://preprocessed-connectomes-project.org/abide/ (accessed August 10, 2023). The rs-fMRI data set[17] used in Section 4 is freely available at https://openneuro.org/datasets/ds000228/versions/1.1.0 (accessed August 10, 2023).

# References

1. Brugere I, Gallagher B, Berger-Wolf TY. Network structure inference, a survey: Motivations, methods, and applications. *ACM Comput Surv*. 2018;51(2):1-39.
2. Simpson SL, Bowman FD, Laurienti PJ. Analyzing complex functional brain networks: fusing statistics and network science to understand the brain. *Stat Surv*. 2013;7:1.
3. Adamovich T, Zakharov I, Tabueva A, Malykh S. The thresholding problem and variability in the EEG graph network parameters. *Sci Rep*. 2022;12(1):18659.
4. Garrison KA, Scheinost D, Finn ES, Shen X, Constable RT. The (in) stability of functional brain network measures across thresholds. *Neuroimage*. 2015;118:651-661.
5. Civier O, Smith RE, Yeh C-H, Connelly A, Calamante F. Is removal of weak connections necessary for graph-theoretical analysis of dense weighted structural connectomes from diffusion MRI? *Neuroimage*. 2019;194:68-81.
6. Heinze G, Boulesteix AL, Kammer M, Morris TP, White IR. Phases of methodological research in biostatistics—Building the evidence base for new methods. *Biom J*. 2022:2200222.
7. Gregorich M, Melograna F, Sunqvist M, Michiels S, Van Steen K, Heinze G. Individual-specific networks for prediction modelling–a scoping review of methods. *BMC Med Res Methodol*. 2022;22(1):62.
8. Sauerbrei W, Perperoglou A, Schmid M, Abrahamowicz M, Becher H, Binder H, Dunkler D, Harrell FE, Royston P, Heinze G. State of the art in selection of variables and functional forms in multivariable analysis—outstanding issues. *Diagn Progn Res*. 2020;4(1):1-18.
9. Sylvestre MP, Abrahamowicz M. Flexible modeling of the cumulative effects of time-dependent exposures on the hazard. *Stat Med*. 2009;28(27):3437-3453.
10. Ramsay J, Silverman B. Functional data analysis. 2nd Springer. *New York*. 2005;
11. Vaughan LK, Divers J, Padilla MA, Redden DT, Tiwari HK, Pomp D, Allison DB. The use of plasmodes as a supplement to simulations: a simple example evaluating individual admixture estimation methodologies. *Comput Stat Data Anal*. 2009;53(5):1755-1766.
12. Di Martino A, Yan C-G, Li Q, Denio E, Castellanos FX, Alaerts K, Anderson JS, Assaf M, Bookheimer SY, Dapretto M. The autism brain imaging data exchange: towards a large-scale evaluation of the intrinsic brain architecture in autism. *Mol Psychiatry*. 2014;19(6):659-667.
13. Craddock C, Benhajali Y, Chu C, Chouinard F, Evans A, Jakab A, Khundrakpam BS, Lewis JD, Li Q, Milham M. The neuro bureau preprocessing initiative: open sharing of preprocessed neuroimaging data and derivatives. *Front Neuroinform*. 2013;7:27.
14. Yu M, Linn KA, Cook PA, Phillips ML, McInnis M, Fava M, Trivedi MH, Weissman MM, Shinohara RT, Sheline YI. Statistical harmonization corrects site effects in functional connectivity measurements from multi-site fMRI data. *Hum Brain Mapp*. 2018;39(11):4213-4227.
15. Fortin J-P, Parker D, Tunç B, Watanabe T, Elliott MA, Ruparel K, Roalf DR, Satterthwaite TD, Gur RC, Gur RE. Harmonization of multi-site diffusion tensor imaging data. *Neuroimage*. 2017;161:149-170.
16. Hardin J, Garcia SR, Golan D. A method for generating realistic correlation matrices. *Ann Appl Stat*. 2013:1733-1762.
17. Richardson H, Lisandrelli G, Riobueno-Naylor A, Saxe R. Development of the social brain from age three to twelve years. *Nat Commun*. 2018;9(1):1027.
18. Šinkovec H, Heinze G, Blagus R, Geroldinger A. To tune or not to tune, a case study of ridge logistic regression in small or sparse datasets. *BMC Med Res Methodol*. 2021;21:1-15.

# Supplementary Material

Flexible parametrization of graph-theoretical features from individual-specific networks for prediction

Mariella Gregorich[1], Sean L. Simpson[2] and Georg Heinze[1*]

[1]Medical University of Vienna, Center for Medical Data Science, Institute of Clinical Biometrics, Vienna, Austria

[2]Department of Biostatistics and Data Science, Wake Forest University School of Medicine, Winston-Salem, NC, USA

*Corresponding author: georg.heinze@meduniwien.ac.at

## Contents

# 1. Simulation design

## 1.1. Data

### 1.1.1. Study design and pre-processing

The Preprocessed Connectomes Project (PCP) provides pre-processed fMRI neuroimaging data from the Autism Brain Imaging Data Exchange (ABIDE). Multiple strategies and pipelines were conducted for the functional pre-processing due to the lack of consensus on the best approach. Following Di Martino et al. [1], we chose the Configurable Pipeline for the Analysis of Connectomes (CPAC) for basic processing with global signal correction and band-pass filtering (0.01 - 0.1 Hz) the preprocessing of the resting-state fMRI data. Time series from region of interests were extracted for the Automated Anatomical Labelling (AAL) atlas [2].

The data of the ABIDE Preprocessed initiative comprises 1112 observations from 539 individuals with Autism spectrum disorder (ASD) and 573 age-matched typical controls (TC) from seventeen sites [3]. The full cohort was restricted according to the quality assurance and sample selection specified in Di Martino et al. [1], which includes, inter alia, the restriction to only males participants due to the low number of females (N=164), yielding N=704. In addition, we omitted 18 observations with missing or unmeasured time series.

*Supplementary Table 1. Baseline characteristics of the full cohort available from the ABIDE Preprocessed initiative and the selected subset of the ABIDE project*

| Variable | Full Cohort |
|---|---|
| Sample size | 1112 |
| Age, years | 17.1 ± 8.0 |
| Handedness, % | |
|   Left | 99 ( 9.1) |
|   Right | 969 (88.8) |
|   Left -> Right | 1 ( 0.1) |
|   Mixed | 6 ( 0.5) |
|   Ambidextrous | 16 ( 1.5) |
| Medication intake, % | 136 (16.8) |
| Diagnostic group, % | 539 (48.5) |
| DSM-IV-TR Diagnostic Category, % | |
|   Autism | 347 (33.4) |
|   Asperger's | 93 ( 8.9) |
|   PDD-NOS | 36 ( 3.5) |
|   Asperger's or PD-NOS | 6 ( 0.6) |
|   Control | 558 (53.7) |
| Full-scale IQ | 108.4 ± 15.1 |
| Verbal IQ | 107.8 ± 16.2 |
| Performance IQ | 106.6 ± 15.3 |

1.1.2. Graph-theoretical initial data analysis

The distribution of edge weights in fMRI connectivity networks is usually centered around zero. Hence, if only positive edge weights are considered and negative edge weights set to zero, the distribution of edge weights tends to follow a a log-normal distribution, with a few strong connections and many weak connections. However, there is also variability in the edge weight distribution across individuals. Some individuals may have more connections with high edge weights, while others may have more connections with low edge weights. This variability may be related to individual differences in brain function or structure, as well as differences in the task or condition being studied; hence we can see variations in the individual edge weight distributions of a randomly selected subset (N=300) of the ABIDE cohort in Supplementary Figure 1.

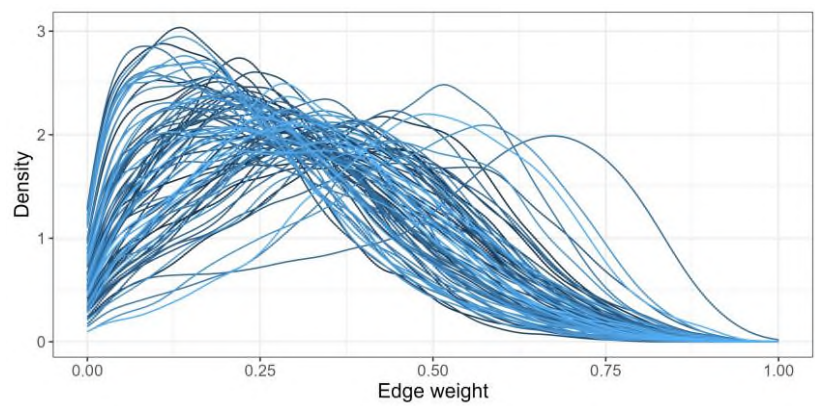

*Supplementary Figure 1. Individual-specific variability across edge weight distributions for a random selected subset of 300 individuals in the ABIDE cohort. Individual edge weight distributions are coloured in blue variations.*

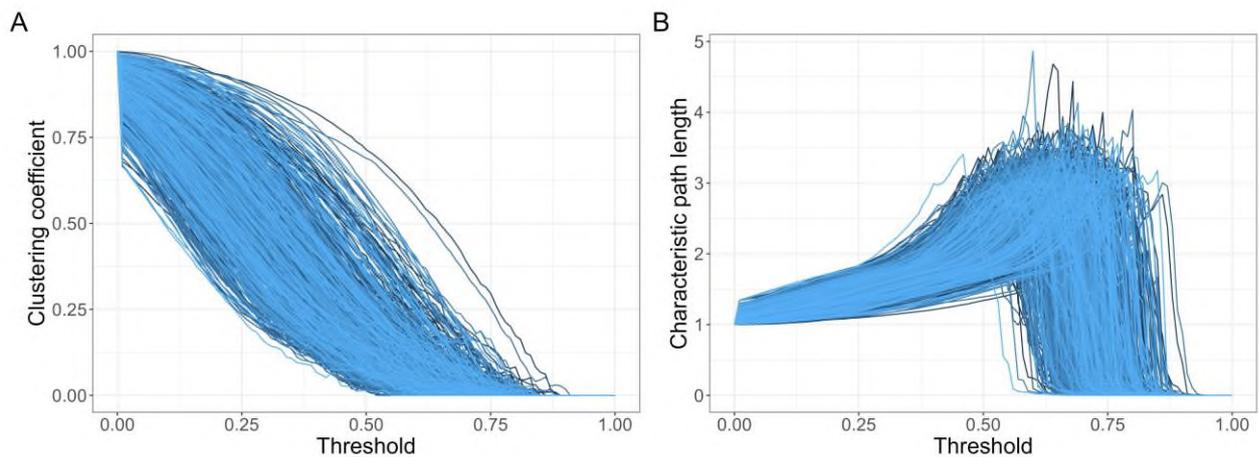

*Supplementary Figure 2. Distribution of the graph theoretical features, clustering coefficient (A) and characteristic path length (B), across varying threshold for weight-based sparsification thresholds for each individual depicted in blue variations.*

## 1.2. Data-generating mechanism

Since we assume that the outcome is generated by a linear regression model

$$y_i = \beta_0 + \beta_1 \sum_{t \in T} \omega(t) x_i(t),$$

We assessed varying forms of the outcome-generating function $\omega(t)$:

(i) a universal threshold for all individuals $i = 1, \dots, N$ as illustrated in Suppl. Figure 3 (left),
(ii) randomly selected thresholds for each individual in Suppl. Figure 3 (right)

and three truly functional forms:

(iii) a flat form as in Suppl. Figure 4 (left),
(iv) an early peak form (Suppl- Figure 4, middle) and
(v) an arc form (Suppl. Figure 4, right).

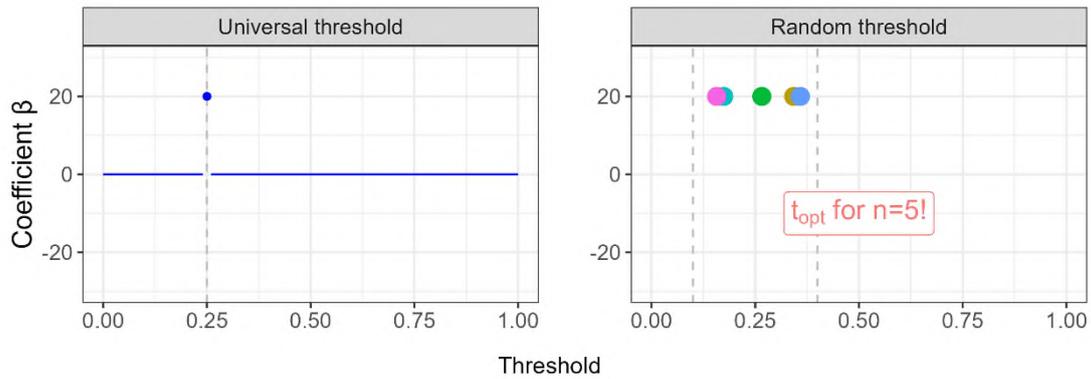

*Supplementary Figure 3. True coefficient of the universal threshold approach and the individual-specific threshold randomly selected from a uniform distribution $U(0.1, 0.4)$.*

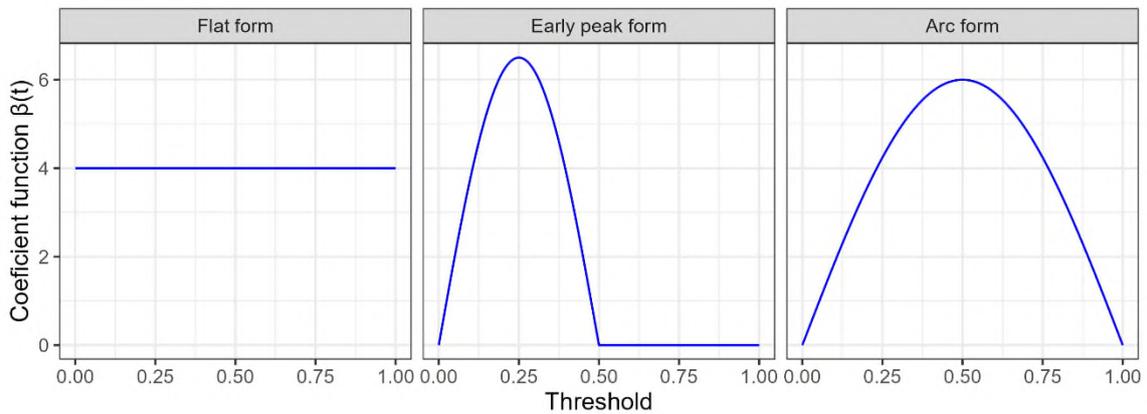

*Supplementary Figure 4. The true functional forms $\beta(t)$ of the OGMs flat, early peak, and arc.*

### 1.2.1. Specification of residual variance

The residual variance $\sigma^2$ in each of the functional form settings (universal, random, flat, half-arc, and arc) was chosen to match the desired levels of $R^2$ (1, 0.6, or 0.3) in the different OGMs. These levels represent no, moderate, and high residual variance, respectively. The specific values for $\sigma^2$ were determined through a preliminary pilot study conducted before the simulation study.

*Supplementary Table 2. Residual variance for the 5 functional forms.*

|  | Residual variance | | |
| --- | --- | --- | --- |
|  | None ($R^2$=1) | Moderate ($R^2$=0.6) | High ($R^2$=0.3) |
| **Functional form** | | | |
| universal | 0 | 2.50 | 5.00 |
| random | 0 | 3.00 | 6.00 |
| flat | 0 | 2.50 | 5.00 |
| early peak | 0 | 2.50 | 5.00 |
| arc | 0 | 2.50 | 5.00 |

### 1.2.2. Contamination of the edge weights

Hardin et al. [4] proposed an algorithm to incorporate noise into a given correlation matrix in a realistic manner such that it remains positive-definite and the condition number is bounded from above. The algorithm by Hardin et al. [4] takes a prespecified $p \times p$ dimensional correlation matrix $\Sigma$, a maximum contamination strength $s$ and a positive integer $m$, and then samples $p$ vectors $\boldsymbol{u_i} \in \mathbb{R}^m$, such that we obtain the $m \times p$ matrix $\boldsymbol{U}$ of the normalized vectors given by

$$\boldsymbol{U} = \left( \frac{\boldsymbol{u_1}}{\sqrt{\boldsymbol{u_1'u_1}}}, \dots, \frac{\boldsymbol{u_p}}{\sqrt{\boldsymbol{u_p'u_p}}} \right) \quad (1)$$

The contaminated correlation matrix $S$ is then given by

$$\boldsymbol{S} = \boldsymbol{\Sigma} + \alpha \delta_i (\boldsymbol{U'U} - \boldsymbol{I_p}) \quad (2)$$

where $\boldsymbol{I_p}$ is the $p \times p$ dimensional identity matrix and the parameter $\delta_i = 1$ in the original algorithm by Hardin. The transformation only affects the off-diagonal elements of $\Sigma$ and ensures that $|S_{ij} - \Sigma_{ij}| \leq \alpha$. The $n$-dimensional vector $\boldsymbol{\delta} \sim U(0,1)$ controls the impact of contamination across the individual-specific networks. The parameter $\alpha$ is specified to reflect a realistic amount of noise in the edge weights. Specifically, for moderate contamination, $\alpha$ is set to 0.15, while in highly contamination scenarios $\alpha$ is increased to 0.3. An example of the impact of the contamination is illustrated in Supp. Figure 5.

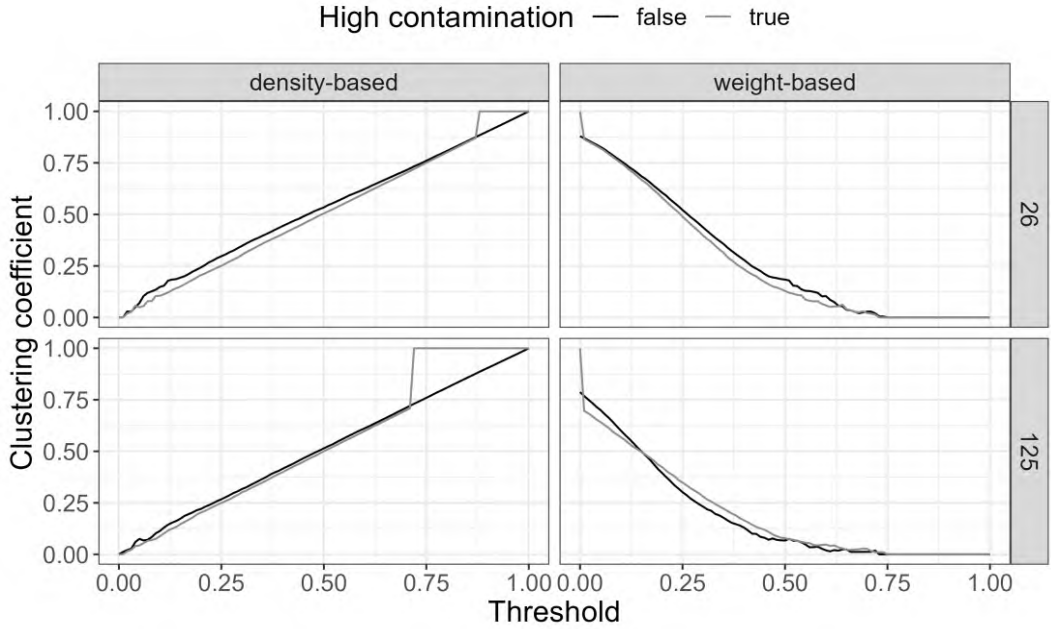

*Supplementary Figure 5. Example of high contamination ($\alpha = 0.3$) for two randomly selected individuals from the ABIDE study cohort.*

### 1.2.3. Monte Carlo error for RMSPE

Because of the substantial computational time required for performing extensive matrix computations, which include repeated thresholding and the calculation of graph-theoretical features on large datasets, the simulation replicates were constrained to $n_{sim} = 500$. Nevertheless, the choice of employing 500 replicates was thoroughly evaluated across all 360 data-generating mechanisms and model construction methods as outlined in [5]. This assessment involved computing the Monte Carlo error for the RMSPE and resulted in a maximum error of 0.08.

### 1.2.4. Additional results

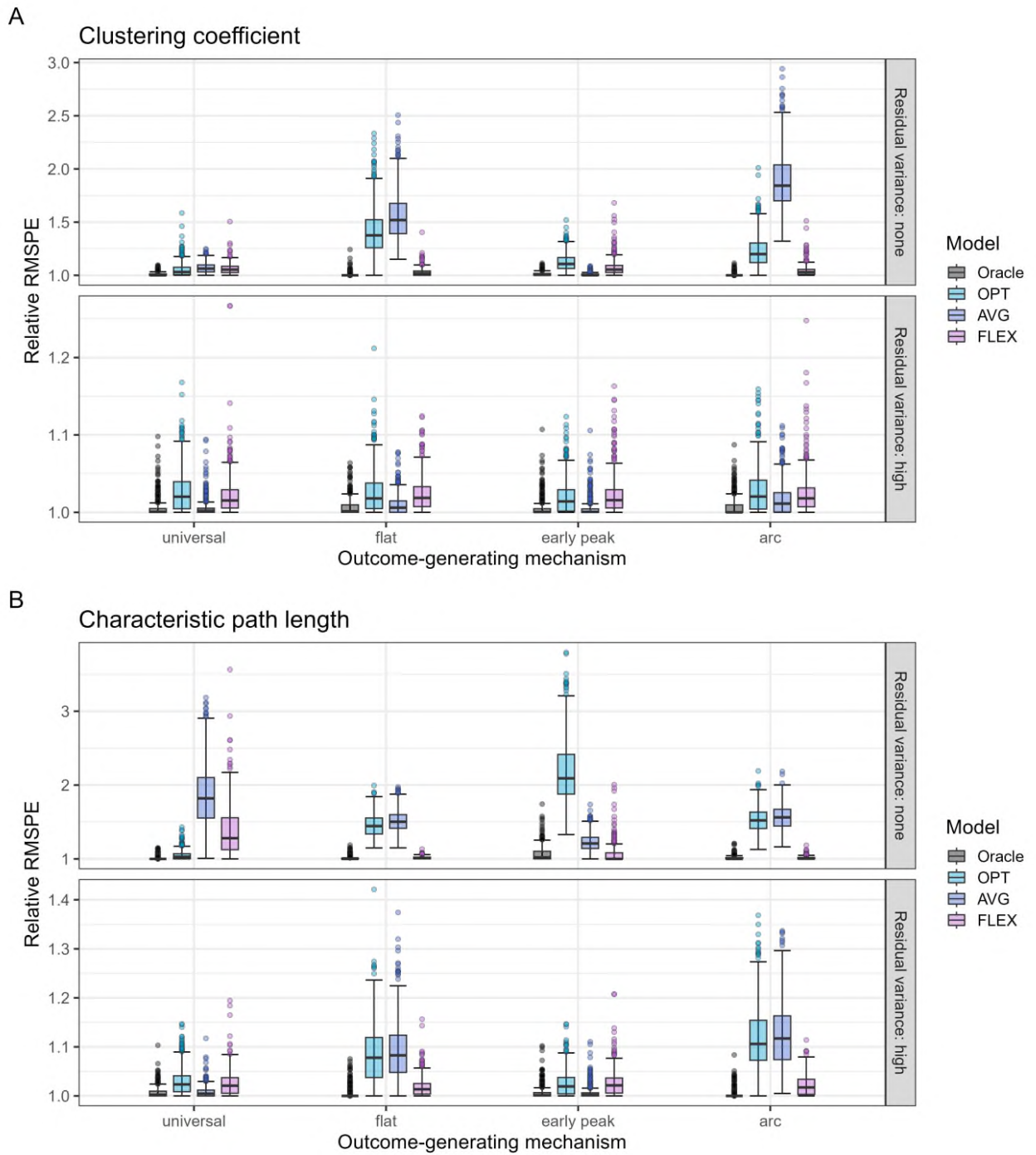

*Supplementary Figure 6. Relative RMSPE of the competitor models and the oracle model for the OGMs universal, flat, early peak and arc for a fixed sample size of n=75 for no residual variance ($R^2=1$) and high residual variance ($R^2=0.3$) and the inclusion of the explanatory variable clustering coefficient (A) and characteristic path length (B) under weight-based thresholding.*

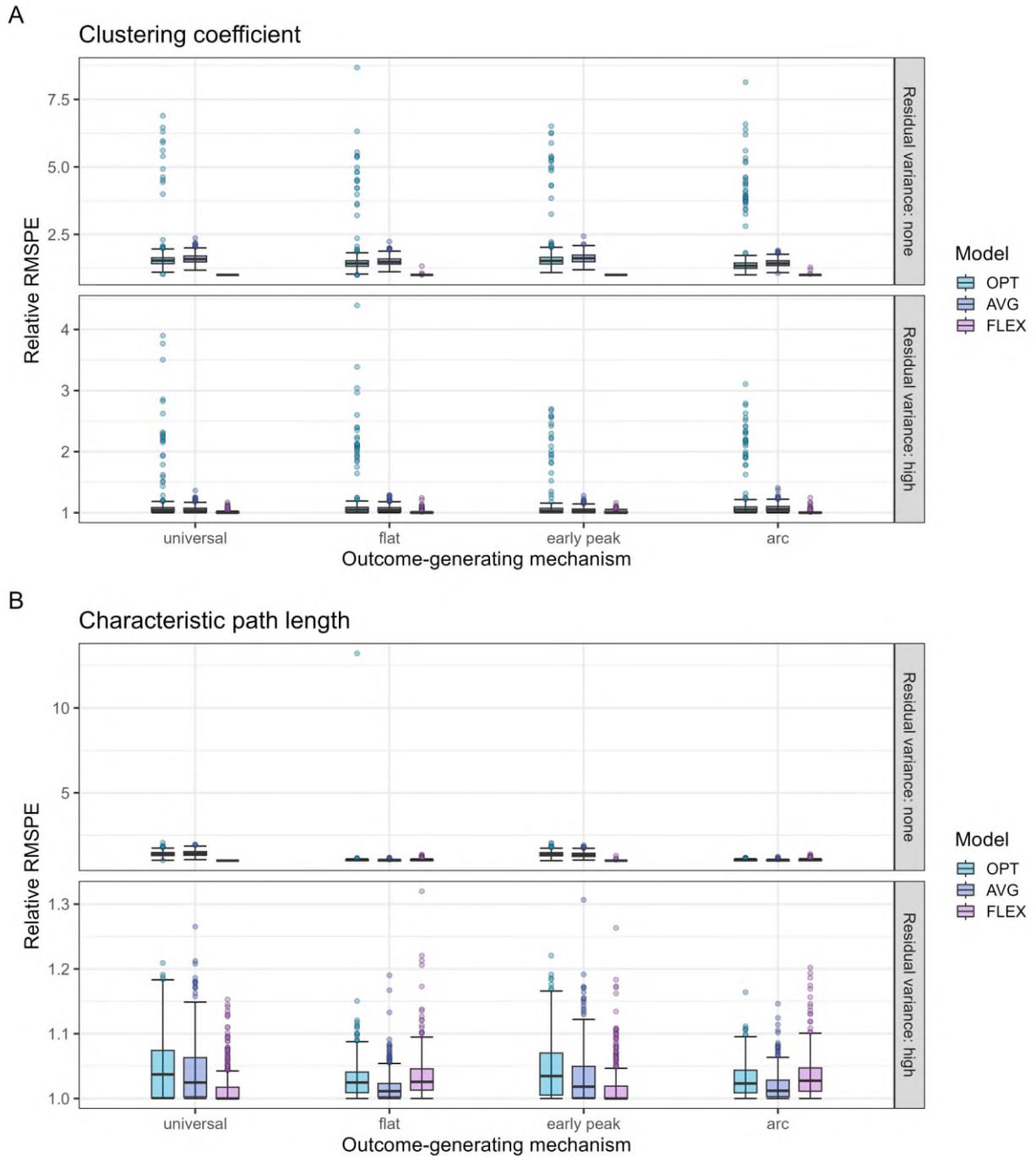

*Supplementary Figure 7. Relative RMSPE of the competitor models and the oracle model for the OGMs universal, flat, early peak and arc for a fixed sample size of n=75 for no residual variance ($R^2=1$) and high residual variance ($R^2=0.3$) and the inclusion of the explanatory variable (A) clustering coefficient and (B) characteristic path length under density-based thresholding.*

*Supplementary Table 3. Calibration slope of the competitor models and the oracle model for the OGMs universal, flat, early peak and arc for a fixed sample size of n=75 for no residual variance (R2=1) and high residual variance (R2=0.3) and the inclusion of the explanatory variable (A) clustering coefficient and (B) characteristic path length under density-based thresholding.*

| Feature | Model | Residual variance | Outcome-generating mechanism | | | |
|---|---|---|---|---|---|---|
| | | | universal | flat | early peak | arc |
| **CC** | Oracle | none | 1.003 (0.995, 1.014) | 1.003 (0.993, 1.015) | 1.003 (0.994, 1.015) | 1.004 (0.992, 1.018) |
| **CC** | OPT | none | 1.002 (0.99, 1.014) | 1.002 (0.981, 1.028) | 1.002 (0.989, 1.016) | 1.002 (0.982, 1.024) |
| **CC** | AVG | none | 1.005 (0.995, 1.018) | 0.998 (0.982, 1.014) | 1.003 (0.994, 1.013) | 0.996 (0.973, 1.017) |
| **CC** | FLEX | none | 1.000 (0.988, 1.013) | 1.002 (0.99, 1.016) | 0.999 (0.986, 1.013) | 1.001 (0.984, 1.018) |
| **CC** | Oracle | high | 1.060 (0.927, 1.226) | 1.051 (0.909, 1.216) | 1.094 (0.935, 1.317) | 1.043 (0.916, 1.203) |
| **CC** | OPT | high | 1.050 (0.864, 1.314) | 1.041 (0.844, 1.292) | 1.086 (0.829, 1.421) | 1.038 (0.855, 1.245) |
| **CC** | AVG | high | 1.062 (0.925, 1.24) | 1.046 (0.918, 1.195) | 1.094 (0.932, 1.312) | 1.034 (0.923, 1.188) |
| **CC** | FLEX | high | 0.992 (0.812, 1.188) | 0.984 (0.788, 1.161) | 0.978 (0.733, 1.227) | 0.991 (0.814, 1.160) |
| **CPL** | Oracle | none | 1.001 (0.997, 1.006) | 1.007 (0.934, 1.088) | 1.002 (0.996, 1.009) | 1.004 (0.935, 1.081) |
| **CPL** | OPT | none | 1.001 (0.994, 1.007) | 0.966 (-1.934, 2.529) | 1.001 (0.984, 1.026) | 0.704 (-2.257, 4.23) |
| **CPL** | AVG | none | 1.006 (0.996, 1.020) | 2.335 (-9.694, 8.146) | 1.002 (0.995, 1.009) | 0.215 (-13.83, 13.913) |
| **CPL** | FLEX | none | 0.998 (0.979, 1.014) | 0.978 (0.886, 1.064) | 1 (0.991, 1.01) | 0.992 (0.909, 1.076) |
| **CPL** | Oracle | high | 1.051 (0.937, 1.216) | 1.270 (0.878, 1.606) | 1.07 (0.938, 1.268) | 1.096 (0.903, 1.339) |
| **CPL** | OPT | high | 1.042 (0.852, 1.313) | -2.003 (-5.836, 4.174) | 1.056 (0.857, 1.346) | 6.716 (-4.449, 6.164) |
| **CPL** | AVG | high | 1.058 (0.932, 1.251) | 0.057 (-15.343, 14.29) | 1.069 (0.941, 1.256) | 12.35 (-13.397, 11.836) |
| **CPL** | FLEX | high | 0.973 (0.791, 1.131) | 0.954 (0.624, 1.287) | 0.979 (0.788, 1.157) | 0.96 (0.658, 1.194) |

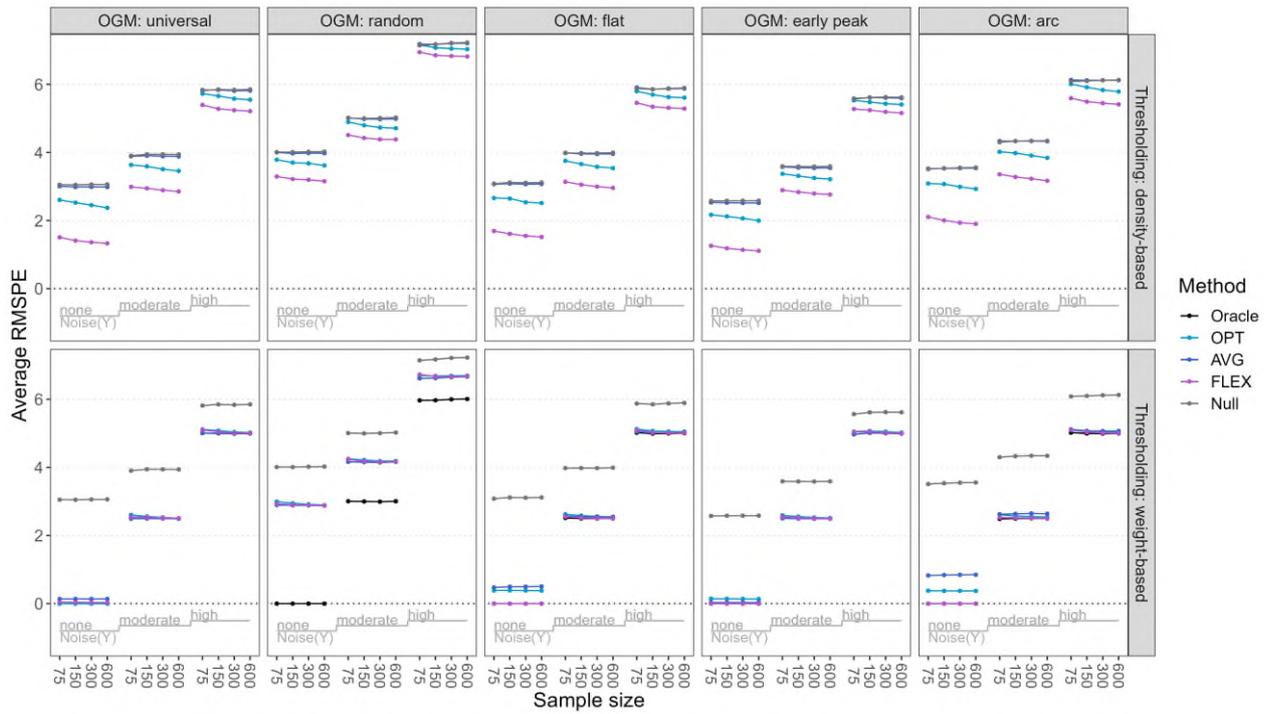

*Supplementary Figure 8. Results across OGMs and varying levels of outcome contamination (none, moderate, high). Clustering coefficient used as graph-theoretical feature and no contamination for the graphs were specified.*

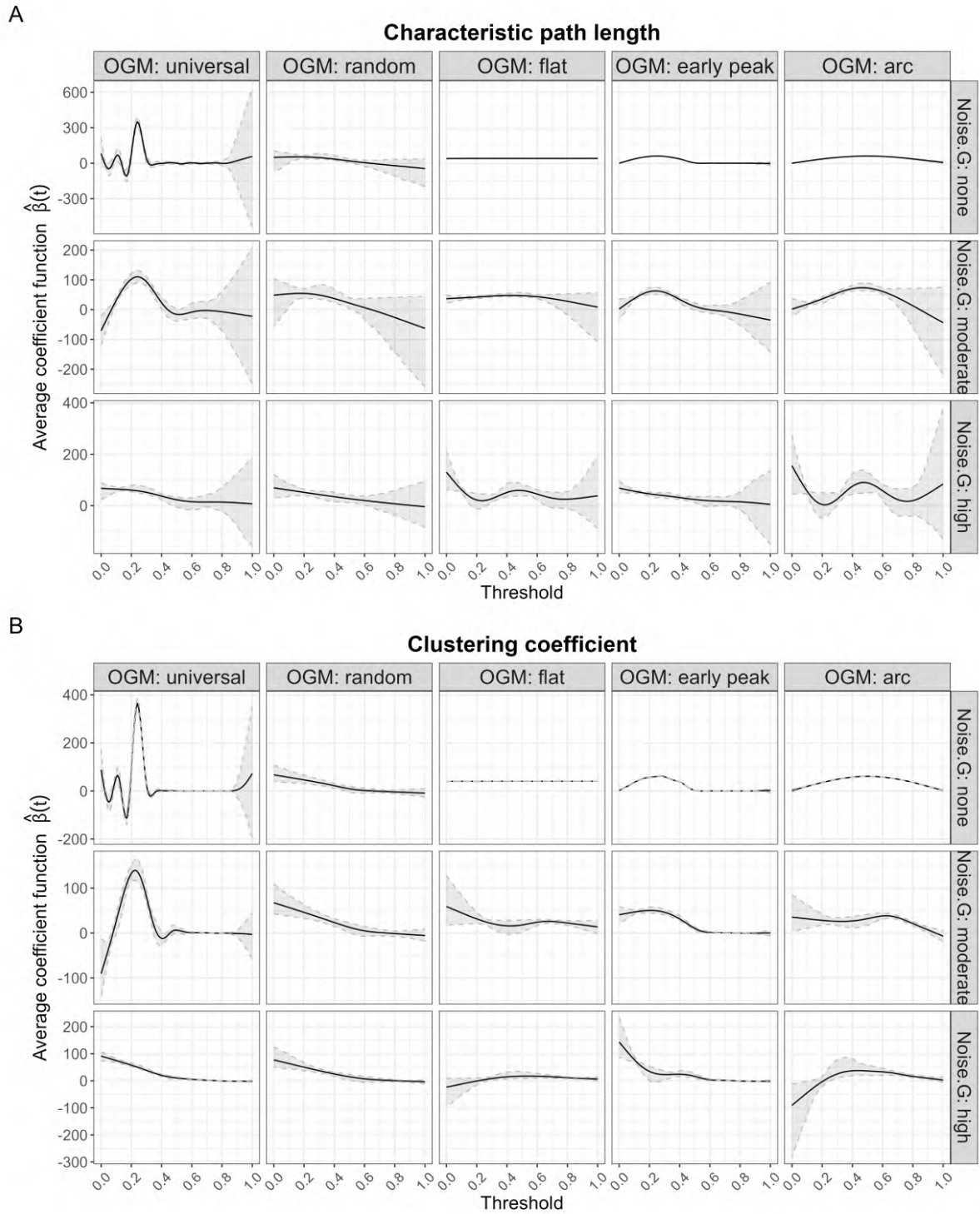

*Supplementary Figure 9. Average functional form $\hat{\beta}(t)$ obtained by the estimated flexible parametrization using penalized splines with 25 segments for the variable clustering coefficient (top) and the characteristic path length (bottom) for varying levels of edge weight contamination across all OGMs. The sample size was fixed at n=75. Weighted-based thresholding was performed.*

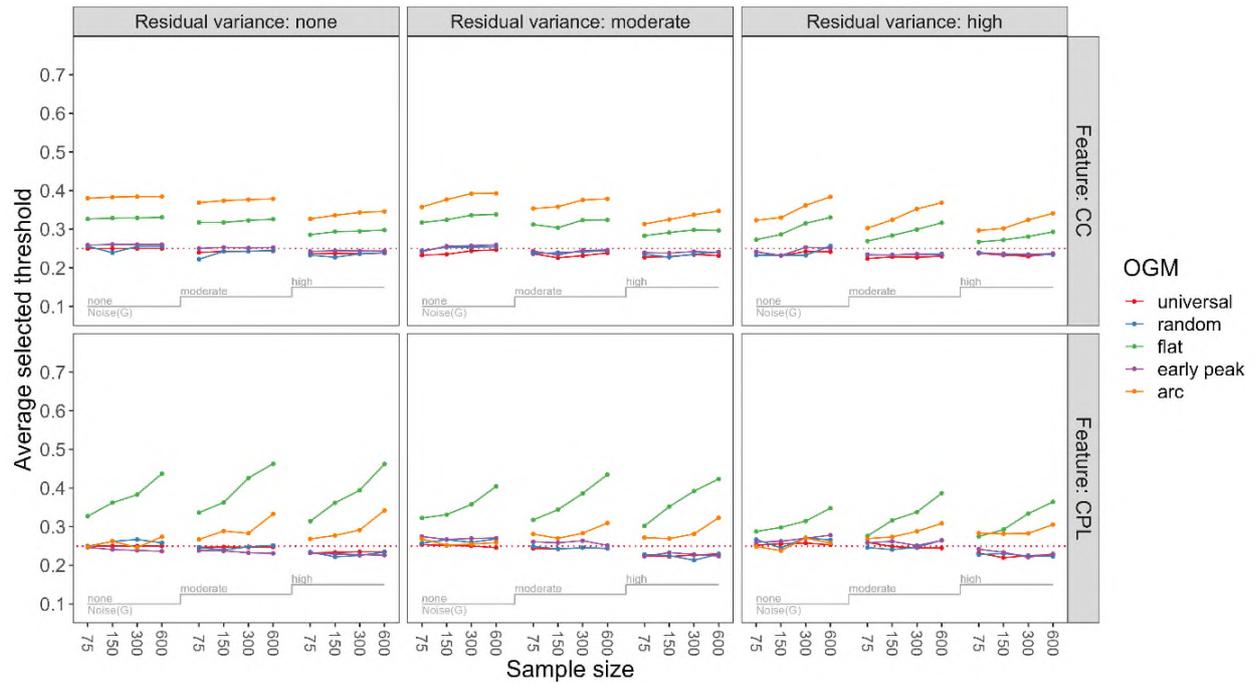

*Supplementary Figure 10. Threshold selection of the OPT model for all OGMs averaged across simulation runs when performing weight-based analysis. The dashed line indicates the true underlying threshold implemented for the OGM 'universal'.*

## 2. Real data example

### 2.1. Study design and pre-processing

The data was obtained from OpenNeuro (as ds000228) available as derivative data that was pre-processed using fMRIPrep. In this experiment conducted by [6], the participants were asked to lie still in the scanner and watch the movie "Partly Cloudy" by Disney Pixar. They were not given any specific task during this time. To begin the experiment, there was a 10-second period of rest with a black screen. Following the rest period, the movie started, and the first 10 seconds consisted of the opening credits, including the Disney castle and Pixar logo. The total length of the experiment was 5.6 minutes. For the atlas choice functional (BASC multiscale) was used. Among the findings of the study, the authors state that functional maturity of each network was related to increasingly anti-correlated responses between the networks. The characteristics of the study cohort are presented in Supplementary Table 4.

*Supplementary Table 4. Demographic characteristics of the study participants*

| Variable | Full cohort (N=155) | Restricted cohort (N=122) |
|---|---|---|
| Age, in years | 7.68 [5.30, 10.97] | 5.98 [4.90, 8.40] |
| Sex, female (%) | 84 (54.2) | 64 ( 52.5) |
| Age $\geq$ 18 years (%) | 33 (21.3) | - |
| Handedness (%) | | |
| Left-handed | 10 (6.5) | 9 ( 7.4) |
| Right-handed | 142 (91.6) | 110 ( 90.2) |
| Ambidextrous | 3 (1.9) | 3 ( 2.5) |

Out of the 155 study participants, not all were children but 33 were already over the age of 18. The oldest study participant was 39 years old, while the youngest was 3.5 years.

## 2.2. Graph-theoretical initial data analysis

For the analysis of individual-specific networks, we used the absolute correlation coefficients to focus on the strength of relationships between variables, disregarding the direction of the correlations. The change in the initial adjacency by disregarding the direction of the correlation is illustrated in Suppl. Figure 11.

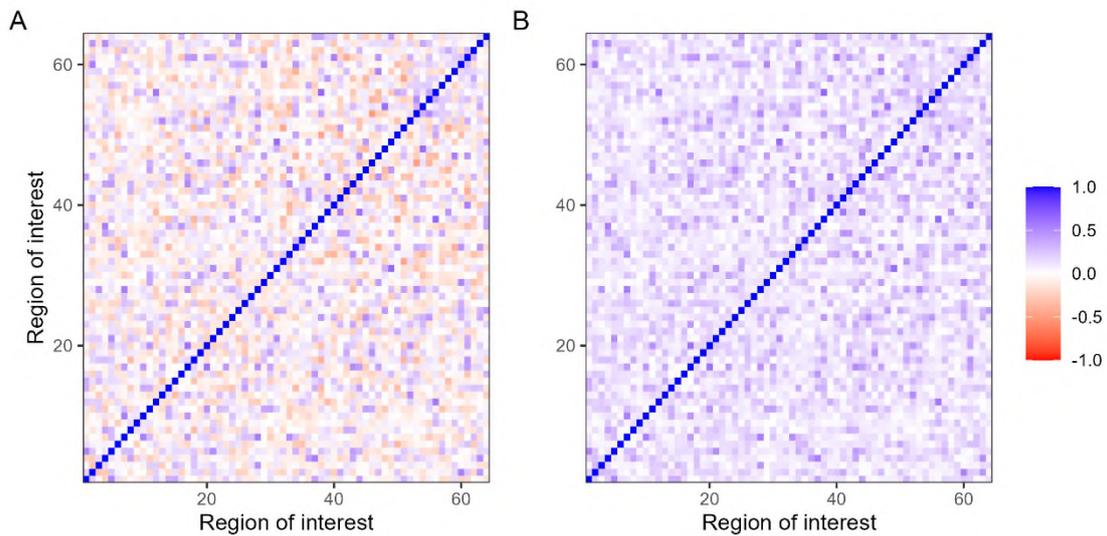

*Supplementary Figure 11. (A) The initial correlation-based adjacency matrix of a randomly chosen individual featuring a network density of 100%, and (B) the corresponding matrix after the transformation to absolute values, with a density of less than 100%.*

The density distribution of the absolute edge weights of each individual-specific network depicted in Suppl. Figure 12 shows the characteristic shape of individual-specific networks inferred by correlation-based network construction with a high number of low edge weights and few strong connections.

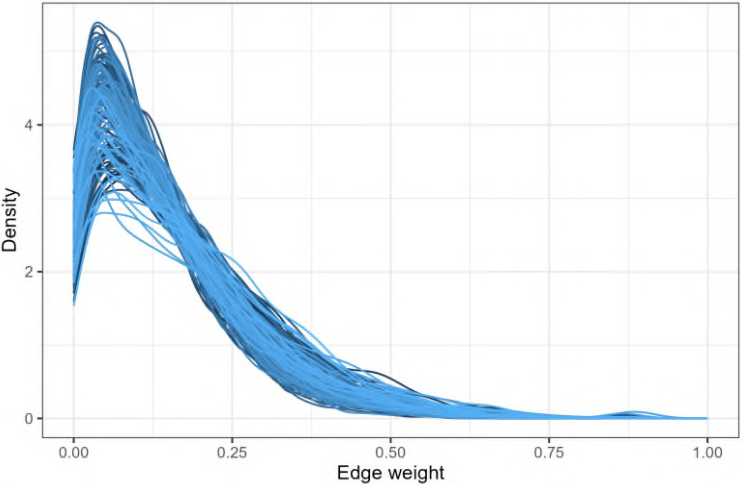

*Supplementary Figure 12. Edge weight distribution of the individual-specific networks*

In Suppl. Figure 13, the individual-specific curves of the graph-theoretical features for increasing threshold levels are depicted and show that these mainly differ by amplitude due to heterogeneity of edge weights across individual-specific networks.

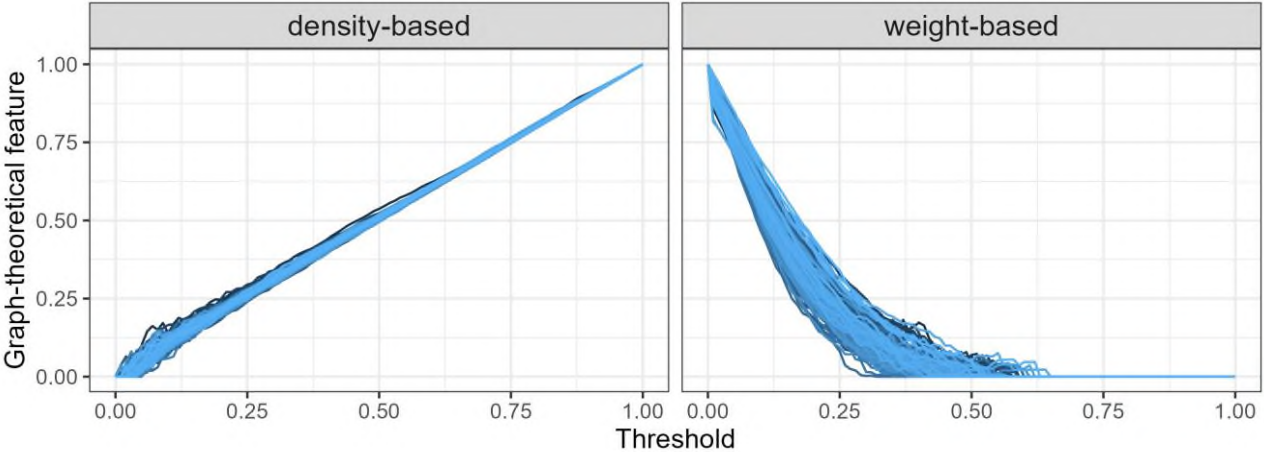

Supplementary Figure 13. *Distribution of the graph-theoretical feature clustering coefficient across the range of thresholds for the individual-specific networks.*

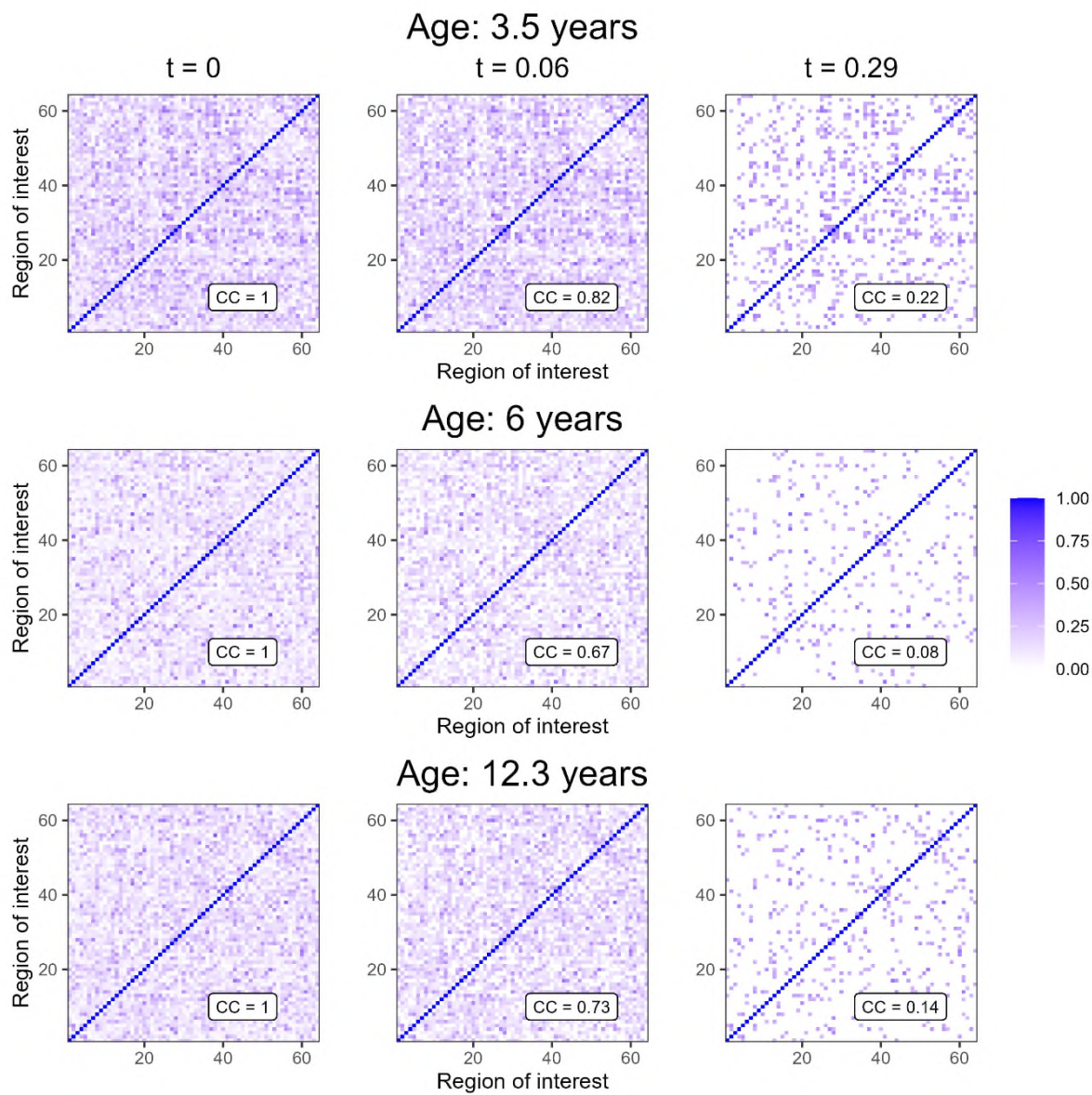

*Supplementary Figure 14. Illustrative adjacency matrices showcasing three individuals from the cohort, representing the minimum, median, and maximum age, using thresholds of 0, 0.06, and 0.29. The presented thresholds align with the peaks in the functional form obtained in the weight-based analysis in Figure 3.*

# References

1. Di Martino A, Yan C-G, Li Q, Denio E, Castellanos FX, Alaerts K, Anderson JS, Assaf M, Bookheimer SY, Dapretto M. The autism brain imaging data exchange: towards a large-scale evaluation of the intrinsic brain architecture in autism. *Mol Psychiatry*. 2014;19(6):659-667.
2. Tzourio-Mazoyer N, Landeau B, Papathanassiou D, Crivello F, Etard O, Delcroix N, Mazoyer B, Joliot M. Automated anatomical labeling of activations in SPM using a macroscopic anatomical parcellation of the MNI MRI single-subject brain. *Neuroimage*. 2002;15(1):273-289.
3. Craddock C, Benhajali Y, Chu C, Chouinard F, Evans A, Jakab A, Khundrakpam BS, Lewis JD, Li Q, Milham M. The neuro bureau preprocessing initiative: open sharing of preprocessed neuroimaging data and derivatives. *Front Neuroinform*. 2013;7:27.
4. Hardin J, Garcia SR, Golan D. A method for generating realistic correlation matrices. *Ann Appl Stat*. 2013:1733-1762.
5. Morris TP, White IR, Crowther MJ. Using simulation studies to evaluate statistical methods. *Stat Med*. 2019;38(11):2074-2102.
6. Richardson H, Lisandrelli G, Riobueno-Naylor A, Saxe R. Development of the social brain from age three to twelve years. *Nat Commun*. 2018;9(1):1027.